\documentstyle[10pt,aasms]{article}
\journalid{337}{15 January 1989}
\articleid{11}{14}
\begin{document}

\begin{center}

     LIMITS FROM {\em CGRO\/}/EGRET DATA ON THE USE OF ANTIMATTER AS A POWER 
              SOURCE BY EXTRATERRESTRIAL CIVILIZATIONS \\

          Michael J. Harris \\
          Universities Space Research Association \\
          300 D Street S.W., Suite 801 \\
          Washington DC  20024 \\
          U.S.A.      \\

~ 

~ \\
~ \\
\end{center}

I argue that the existence of cold antimatter in bulk is not permitted by the 
Standard Model, so that if a $\gamma$-ray signature from antiproton 
annihilation 
were to be detected, it must represent either new physics or the action of 
intelligence.  Time variability of the signal would strongly support the 
second alternative.  The entire sky was scanned at the relevant energies 
(30 -- 928 MeV) by the EGRET experiment on board the {\em Compton\/} Gamma Ray 
Observatory during 1991--1995. A search of this database for the antiproton 
annihilation signature yielded only upper limits on the flux (an intriguing 
spectrum detected from QSO 2206+650 = 3EG J2206+6602 
is probably not related to SETI).  The 
all-sky, longterm 99\% upper limit is $2.3 \times 10^{-8}$ photon/(cm$^{2}$ 
s); it is a 
factor 10 worse in the Galactic plane due to the higher diffuse $\gamma$-ray 
background emission.  I give brief, but quantitative, illustrations of what 
this limit means for nearby intelligent activities.

\clearpage

\section {Introduction}

It is a quite general consequence of extensions of the Standard Model of
particle physics, such as supersymmetry, that the early Universe was 
characterised by a small excess of baryons over antibaryons [1].  At some point
in the cooling of the Big Bang their mutual annihilation produced the enormous 
photon flux ($\sim 10^{9}$ photons for each of the few surviving baryons) 
which we see 
today as the cosmic microwave background.  But regardless of how the baryon 
excess originated, ordinary Standard Model physics requires that the 
annihilation process was extremely efficient [2].  It is thus widely believed 
that no significant numbers of antiprotons ($\overline{p}$), having survived 
from the Big 
Bang, can subsequently have cooled to form substantial "domains" of antimatter, 
recombining with the corresponding antielectrons (i.e. positrons, $e^{+}$).

Any $\overline{p}$ present in the Galaxy today must therefore have been 
created by 
well-understood high-energy processes.  The energies involved are very 
large, given the 
$\overline{p}$ rest-mass $\sim 1$ GeV, and only one astrophysical environment  
is adequate -- the high energy Galactic cosmic ray flux.  The basic physical 
mechanism involved is for a cosmic ray particle with energy 
$\gg 2 m_{p}c^{2}$ incident 
on a proton at rest in the interstellar medium (ISM) to create a shower of 
energetic hadrons which may include $\overline{p}$ and pions (neutral 
$\pi^{0}$, charged $\pi^{+}$, $\pi^{-}$) among others.

The $p\overline{p}$ 
annihilation sites are most germane, since I propose to detect gamma 
rays from the annihilation process.  This may be represented as [3]:
\begin{equation}
 p + \overline{p} \rightarrow 2\pi^{0} +1.5 \pi^{+} +1.5 \pi^{-} + 0.05 K     
\end{equation}
where the multiplicities are average values and K includes all kaon species.  
The 
$\pi^{0}$ decay very rapidly into two $\gamma$-rays, whose energies in the rest 
frame are distributed as shown in Fig. 1 [4].

From the considerations above one may conceive of three $\overline{p}$ 
production processes 
and their annihilation characteristics:

(1) New physics. \\
Khlopov et al. [5] among others have proposed an inflationary
model with nonhomogeneous baryosynthesis 
which produces cold antimatter today in regions of sizes of the order of 
globular clusters ($\sim 10$ pc and $\sim 10^{6} M_{\sun}$).  A detailed
model of the flux, spectrum and Galactic distribution expected from 
annihilation of antimatter ejected from these stars has been 
developed by Golubkov and Khlopov [6], who show that the Galactic halo
contains a substantial steady-state population of $\overline{p}$ fed by
massive stellar winds and supernova ejecta.  My interest here is in their
computed spectrum, which they show to be compatible with EGRET
measurements of the $\gamma$-ray flux from the halo.  The annihilation
spectrum of Fig. 1 will be distorted by the velocities of the stars and
ejecta (and of $\overline{p}$ accelerated by shocks in the ejecta) and is
shown to be much more sharply peaked then Fig. 1. Its distribution is
naturally diffuse and steady.  

(2) "Hot" cosmic ray antiprotons. \\
These $\overline{p}$ annihilate in flight with ISM protons, producing a 
relativistic $\pi^{0}$ 
and then Doppler-shifted $\gamma$-rays.  The spectrum of Fig. 1 is 
kinematically 
broadened and flattened [7].  It must contribute to the steady diffuse 
$\gamma$-ray background from the Galactic plane at some level, but 
$\pi^{0}$ can be produced by other high energy reactions, and
the amount coming from $\overline{p}$ annihilation will be 
swamped by decay of $\pi^{0}$ produced directly by the much more common
$p + p$ interaction.  Even so, the ISM 
$\pi^{0}$-decay
feature is quite weak, and it is 
is necessarily accompanied by a high level of continuum (smooth spectral
power laws) from other high energy proton and electron 
processes, which tends to swamp it (\S 2.2).

(3) An artificial source. \\
The high-energy interactions creating $\overline{p}$ are performed by humans in particle 
accelerators.  It has long been realized that, if practised on a large enough
scale, $p\overline{p}$ 
annihilation could be a very valuable power source for applications 
requiring portability (e.g. rocket propulsion [8]), that the emission of 
$\gamma$-rays is a signature, and that these $\gamma$-rays could be detected 
[9].  The spectrum would be the cold spectrum of Fig. 1, but its other 
characteristics are rather speculative; I will assume here that portability 
implies small size, and so by the usual astronomical argument possible 
variability.  For high-energy gamma ray telescopes, with their bad resolution, 
small size effectively means a point source.  The time-scales on which I 
searched for variability were $\sim 14$ d (\S 2.1).

I conclude that the artificial use of antimatter as a power source has 
unique characteristics which can be used to identify it, i.e. a "cold" 
spectrum of the form of Fig. 1 without a sharp peak but narrower than the
broad Galactic plane $\pi^{0}$ feature, predominance over any accompanying 
continuum, and possibly variability.  The spectroscopic
aspects of this will be discussed further in \S 2. 

The production and cooling of antiprotons by the accelerator technique
is inefficient and extremely costly in energy.
A back-of-the-envelope calculation shows that gamma rays will only be detectable
from an undertaking which is truly immense by human standards (\S 4.1).
However, it is not possible a priori to constrain the nature of the 
extraterrestrial beings in any way whatever.  I do not believe that the 
scale of the phenomenon can be used as an argument against its possibility.
For example, note that processes (1) and (3) above are not incompatible, in 
that antimatter might be "mined" on a large scale from a domain by intelligent 
beings.  In this case one might expect to observe both variable and steady 
annihilation signatures.

Earlier discussions of this topic focused on the role of antimatter in 
propulsion of extraterrestrial spacecraft [9,10,11]; 
Harris [12] also made a search 
for $e^{-}e^{+}$ annihilation powered spacecraft in 
$\gamma$-ray burst data.  In 
this application of antimatter, large Doppler shifts would obviously be another 
distinctive sign of artificiality.  The data to be used here probably do not 
have sufficient energy resolution for this approach to be used, however 
(\S 2.1).

This paper will focus on possibility (3), an artificial source.  The 
evidence for the possibility (1) of new Big Bang physics creating large
antimatter domains is thoroughly explored in Ref. [6] and
will not be addressed here.

\section{Observations and Analysis}

\subsection{Observations}

This paper is based on measurements made by the Energetic Gamma Ray 
Telescope 
(EGRET) on board the {\em Compton\/} Gamma Ray Observatory ({\em CGRO\/}).  
After 
launch into low Earth orbit (LEO) in Spring 1991, the mission was divided into 
Phases of approximately 1 year in length, except for Phase 1 which lasted 18 
months.  The satellite re-entered on 4 June 2000 during Phase 9.  EGRET was 
fixed at the rear of the spacecraft, and could therefore only be pointed by 
manoeuvering the whole spacecraft.  This was done at approximately 2-week 
intervals, each pointing being referred to as a Viewing Period (VP).

The physical principles behind the operation of EGRET [13] are shown in Fig. 2 
in very schematic form.  The key interaction is the formation of a 
particle-antiparticle pair by an incoming $\gamma$-ray photon which has energy 
$>2 m c^{2}$.  By far the easiest particles to produce are $e^{-}$ and 
$e^{+}$, and $mc^{2}$ is thus 
the electron rest mass 0.511 MeV.  In practice the cross section for this does 
not become appreciable until energies of tens of MeV.  The cross section is 
greatly enhanced in the presence of electric charge, which essentially means an
atomic nucleus.

The front end of the telescope therefore consisted of a stack of 27 tantalum 
plates, whose high nuclear charge of 73 enhanced pair creation.  In order to 
actually register a detection ("count"), they were interleaved with digital 
spark chambers, i.e. gas-filled chambers with a high voltage across them which 
suddenly discharge when an $e^{-}$ 
or $e^{+}$ ionizes a gas molecule.  It can be seen 
from Fig. 2 that the sequence of discharges enables the direction of the photon 
to be measured.  The back end consisted of an array of massive NaI scintillator 
crystals (the Total Absorption Shower Counter, or TASC) which were intended to 
stop the $e^{-}$ and $e^{+}$, 
so that the total scintillator light (as measured by 
adjacent photomultiplier tubes) measured the total energy of the photon.

The LEO environment is heavily irradiated by other sources of charged 
particles, such as Galactic cosmic rays, which can seriously compromise 
$\gamma$-ray 
measurements, since at these energies they outnumber them by over 1000 to 
one.  Two features of the instrument were designed to mitigate this.  First, it
was surrounded by an anticoincidence shield, i.e. a large plastic scintillator
which, when it detects a particle from the outside, causes the main instrument 
to be shut off for a very short time ($\sim 100$ ns).  Second, fast 
time-of-flight
electronics between the spark chamber assembly and TASC checked the 
$e^{-}$ and $e^{+}$
trajectories for consistency with an incoming photon before allowing the spark 
chambers to switch on.  The rejection of charged particle backgrounds proved to
be extremely efficient, and they never became a problem for EGRET.

Photons in the energy range $\sim 30$ MeV to 30 GeV can be detected by these 
processes.  Photons with too high energies penetrate the Ta converters without 
interacting.  On the other hand, below 100 MeV the photons tended to be 
deflected (or even absorbed) by the front end components.  This introduced 
errors into EGRET low-energy measurements which could never be fully corrected 
for, and which had an impact upon my analysis (\S 3).  

The energy resolution of EGRET was poor.  The photons were binned into ten 
energy channels which are listed in Table 1.  Only channels 0-8 were used 
in my analysis.  The poor resolution ($> 50$\% above 100 MeV) made 
measurements of Doppler shifts unrealistic.

The accuracy of photon direction measurement (equivalent to point spread 
function [PSF]) varied strongly with photon energy, from several degrees below 
100 MeV to $0.8^{\circ}$ at 1 GeV; for the energies of Fig. 1 the average 
value was about $2.5^{\circ}$.  It also varied with distance from the centre 
of the field of view (FOV), which was $\sim 80^{\circ}$; sensitivity and PSF 
were poor at off-axis distances $>30^{\circ}$.  Angular resolutions actually 
achieved in analysis are of 
course much better if there are enough counts. 

A major constraint on the capabilities of EGRET was the availability of the 
Ne/Ar gas mix used in the spark chambers; being non-renewable, this began 
to run out at the end of Phase 4 (October 1995), after which EGRET was used 
as little as possible.  I have used data from Phases 1--4 in my analysis.

The EGRET instrument team have presented their data in the form of maps for 
each VP, and for the combined Phase 1--4 data, for three quantities: the raw 
count numbers, the $\gamma$-ray 
intensities (i.e. count rate per unit live time, 
per unit area, per unit energy, using an effective area weighted for the 
detection efficiency, PSF and location in the FOV), and the exposure 
(essentially, the variation of the live time across the map).  The maps are on 
a $0.5^{\circ} \times 0.5^{\circ}$ grid.  They are readily available in zipped 
form via FTP; see 
{\em  http://cossc.gsfc.nasa.gov/cossc/EGRET}.  
I searched the Phase 1--4 sum spectra from the entire sky for an annihilation 
signature.

\subsection{Analysis Principles}

When binned into the EGRET energy channels of Table 1, the 
$p\overline{p}$ annihilation 
spectrum of Fig. 1 appears as in Fig. 3.  This strongly suggests that a 
search for that spectrum should begin by selecting map locations with excess 
counts in EGRET channel 4.  Before applying this criterion naively, it is 
necessary to consider possible backgrounds which might affect these 
energies (150-300 MeV).

Instrumental backgrounds are low (\S 2.1).  The relevant background 
here is the entirety of known cosmic $\gamma$-ray sources.  In general this 
means known and measured large-scale diffuse fluxes from the Universe at large 
and from the Galactic halo, plane and solar vicinity.  The measurement of these
was a major objective of the EGRET experiment, but for my analysis they 
were a nuisance.  The salient features of these spectra are the following.

The extragalactic background and halo-plus-local 
(defined as $|b| > 10 ^{\circ}$) spectra were 
separated from each other by [14], who measured the sum of them as a function 
of latitude $b$, 
and extrapolated to $90^{\circ}$ (this is more accurate than any 
single measurement of the extragalctic emission at $90^{\circ}$).  
In this way, the 
extragalactic background spectrum was found to be a power law of index -2.1.  A 
power-law 
index $\sim~ -2$ is observed from 
the halo due to cosmic ray proton and electron processes in the ISM 
[15].  Only the 
combined halo-plus-extragalactic spectrum was needed for my analysis, which I 
assumed to be a power law of index $\sim ~-2$.

The strength of this spectrum increases as $b$ decreases.  This is true in a 
much more marked way of the emission associated with the Galactic plane (the 
component with $|b| < 10^{\circ}$).  Since by far the bulk of the ISM resides 
in the 
plane, this component is very intense, which made it much easier for EGRET to 
measure details of its spectrum.  Thus, superimposed on the power law of index 
$\sim~ - 2$ which comes from typical cosmic ray-ISM interactions EGRET detected 
two features: a broad but weak peak centred at a few hundred MeV, merging 
into a 
general excess of emission at energies above 1 GeV.  The former might be the
spectrum from relativistic $\pi^{0}$ decay (\S 1); the latter is unexplained 
[16,17]. 

Of these various spectral features, which are important at various latitudes, 
only the power laws were a significant background problem for my analysis.  In 
principle the broad $\pi^{0}$ 
decay feature from the ISM in the Galactic plane might 
contaminate any extraterrestrial candidate $p\overline{p}$ spectrum from a 
point source; 
however in practice it is too weak on spatial 
scales $~\sim 1^{\circ}$ relative to
the strong continuum,
and the shape is too dissimilar (it is much broader).  I did not consider
energies significantly above 1 GeV and therefore neglected the alleged excess.
I thus had to take into consideration contamination from a power-law spectrum 
whose intensity rose sharply in and near the Galactic plane.

Some effects of this can be seen in Fig. 4, which shows schematically a 
power law superimposed on the $p\overline{p}$ 
annihilation spectrum.  First, it is 
clear that the annihilation spectrum is completely dominated by the power 
law at energies below 100 MeV (EGRET channels 0--2); therefore channels 3, 4 
and 5 only will play a role in detection.  Second, although Fig. 3 shows 
that channel 4 in general contains the most counts, the 
{\em contrast\/} between 
the annihilation function and the power law is greatest in channel 5.  
Third, the power law itself is not very well characterised; it is 
"anchored" at the high-energy end by channels 6--8, none of which in general 
contains many counts, and at the low-energy end by channels 0--2, which are 
subject to systematic errors (\S 2.1).  

\subsection{Analysis (Details)}

The considerations in \S 2.2 were taken into account in the design of 
the analysis.  My overall strategy was to search the EGRET Phase 1--4 sky 
map point by point for spectra of the form of Fig. 4.  This tacitly assumes 
that the source(s) have a large duty cycle (\S 4).  When a source was 
detected, I went back to the individual EGRET VPs which covered the same region 
of the sky.  A measurement of the same form of spectrum was made in each VP and 
the results were arranged in a time series.  A variable level would 
strengthen the case for an 
candidate extraterrestrial power source (\S 1). 

The process involved several steps:

(1a) The EGRET PSF was crudely approximated by a uniform circle of 
radius $2.5^{\circ}$ centred on each grid point in the EGRET Phase 1--4 map.
The spectral intensities from each point within the circle were averaged 
over.  The counts from each point were summed.  Attention then turned to 
the next point in the grid $0.5^{\circ}$ away, in the manner of Fig. 5.  This 
procedure assumes that the measurement precision is limited by small-number
statistics. \\
\indent (1b) This 
procedure was modified within $8.5^{\circ}$ of the Galactic plane, where 
many counts were registered even within single $0.5^{\circ}$ grid points.  
If several 
sources occur within the PSF circle then measurement is confusion-limited by 
this contamination, and it is better to measure the spectra from the 
individual grid points.  \\
\indent (1c) In both cases an error equal to the square root of the counts was 
assigned 
to each channel.  Typically the cut at $|b| = 8.5^{\circ}$ 
corresponded to a level of 
several tens of counts in channel 4.

(2) An automated search selected those spectra which had a peak at channel 4 
in counts.  Channels 2 and 3 were constrained to be less than channel 4, 
channel 5 was constrained to be less by a factor 2 (Fig. 3).

(3) Many of these spectra turned out to be unusable for various reasons, for 
example zero counts in one or more channels.  The selected spectra were 
examined visually for these problems.  The criterion used was that an excess in
channel 5 be visible, since this channel contrasts best against a power law 
background.

(4) It is obvious from Fig. 5 that most of the spectra generated in step 1a
are not independent measurements, since successive PSF circles overlap.  Away 
from the Galactic 
plane ($|b| > 8.5^{\circ}$) clusters of the selected spectra in 
$\sim 2.5^{\circ}$ groupings were identified, from each of which a single 
representative spectrum
at or near the centre was retained.  A large majority of the selected spectra 
fell into these groupings, which suggests that the selection process was not 
discovering statistical fluctuations in counts.

(5) The surviving candidate spectra were fitted with a spectral 
model consisting 
of the annihilation function plus a power law (as in Fig. 4).  The fits were 
performed by the IDL utility nonlinear least-squares fitting routine 
CURVEFIT.

(6) If a candidate point showed a significant level of 
$p\overline{p}$ annihilation 
emission, the individual VPs contributing to the Phase 1--4 spectrum were 
examined for variability or constancy.  The positions of known EGRET sources 
were excluded as being background, and presumably having been examined 
already [18]. 

Table 2 shows the number of candidates resulting from each of these steps.

\section{Results}

The amplitudes returned by CURVEFIT for the 
$p\overline{p}$ annihilation function represent
the results.  The average results for the two zones $|b| > 8.5^{\circ}$ and 
$|b| < 8.5 ^{\circ}$ are shown in Table 2.  Before going into more detail, a 
comment on the errors is necessary.

The errors are clearly much larger in the Galactic plane.  This is readily 
understood in terms of the background level (\S 2.2).  In terms of Fig. 4, 
the underlying power law is elevated by more than an order of magnitude, and 
the annihilation function becomes invisible against it.  Out of the plane there
is a weaker rising background as the plane is approached, but another source of
error --- poor statistics due to lesser EGRET exposure --- 
dominates in the high-latitude regions where this 
background 
is low.  These two effects work in opposite directions in $b$, so that
above 
$|b| = 8.5^{\circ}$ the error is not a strong function of latitude.  Quoting an 
average value for the error as in Table 2 is therefore meaningful.

The criterion as to whether an annihilation signature has been detected at any 
point is the significance $\sigma$ 
of the measurement over and above the estimated error
(of which average values at 99\% confidence are quoted in Table 2).  Rather than
absolute intensities, I therefore plot in Fig. 6 the distribution of the 
results in standard deviations. If the measurement errors were randomly
distributed, this would in general be a Gaussian of mean 0 and standard 
deviation 1; it is obvious, however, that the strong selection procedures which
I applied in \S 2.3 in order to isolate a positive result have heavily 
biased the distribution towards positive values.  Hence the measured 
distribution of significances is the high-$\sigma$ "tail" of a hypothetical 
distribution of all measurements (including points that were rejected by my 
selection procedure and never measured).  If the errors were randomly 
distributed about a null result this hypothetical distribution would be a
Gaussian distribution of mean zero and standard deviation 1 normalized to the 
number of possible independent EGRET PSFs.  Given the Cartesian $0.5 ^{\circ}$ 
EGRET gridding 
and my approximation of the PSF by a $2.5^{\circ}$ circle there are $\sim 720
\times 360/(\pi \times 5^{2}) = 3300$ independent PSFs.  

The high-$\sigma$ "tail" of 
this expected distribution is overplotted in Fig.6.  
It is clear that there are a significant number of measured values in excess of
what is expected ($\sigma > 2.5$; 
see also Table 2, step 5).  I compared the corresponding spectra with 
typical ones within the expected range of significances; examples of the two 
types are shown in Figs. 7a and 7b.  In almost all cases, the supposedly 
high-$\sigma$ annihilation features were clearly artefacts of anomalous count 
rates in EGRET channels 0--2.  For example, it is clear in Fig. 7b that, if 
there were not an anomalous low flux in channel 0, the spectrum would be well 
fitted by a power law.  Since, as pointed out in \S 2.1, channels 0--2 are 
subject to poorly-understood systematic errors, I conclude that the
measurements of highly significant 
$p\overline{p}$ annihilation fluxes in Fig. 6 are in 
general spurious.

The only spectrum remaining after I had rejected those with problems in 
channels 0--2 was peculiar.  At location $l = 107.25^{\circ}$, $b = 
11.5^{\circ}$, a Phase 1--4
spectrum was obtained which did not show obvious systematics in channels 0--2 
and which showed the annihilation feature at a significance 
$3.8 \sigma$ (a nominal
probability of $\sim 2 \times 10^{-4}$ 
of occurring by chance); however, there was an 
unexpected excess in channel 7 (Fig. 7c).  This is in fact the quasar QSO 
2206+650, which was detected by EGRET (3EG J2206+6602, 
[19]), and the source should probably be 
rejected as an ETI candidate for both of these reasons.  As an illustration of 
the procedure which might in future be applied as an ETI test using variability
(\S 2.3, analysis step 6), I fitted the usual power-law plus $p\overline{p}$ 
model to 
the EGRET spectra from the 9 VPs in which $l = 107.25^{\circ}$, 
$b = 11.5^{\circ}$ was within the FOV.  Interestingly the $p\overline{p}$ 
light curve apparently shows variability 
(Fig. 7d), which would have been a characteristic of ETI activity.

\section{Discussion}

\subsection{Annihilation Flux Limits}

The search performed here is essentially complete and homogeneous in space.
I therefore adopt an upper limit on the steady $p\overline{p}$ annihilation 
flux of 
$2.3 \times 10^{-8}$ photon/(cm$^{2}$ s) from any point outside the 
Galactic plane, with a
limit about a factor 10 higher in the plane.  The upper limit on transient 
fluxes is less easily established.  My search method was designed for steady 
sources or those with a high duty cycle --- i.e., a duty cycle long enough 
that the transient emission dominates the overall Phase 1--4 emission.
Since a short, strong event of low duty cycle (1 VP) will do this, as will
a long, slow, weak event lasting for years, one must choose a "standard" 
duty cycle, and determine the corresponding "standard" flux limit. 

Let us consider the behaviour of the source at $l = 107.25^{\circ}$, $b = 
11.5^{\circ}$ (Fig. 7d) as a test case.  This source 
is known to produce a detectable annihilation flux when
averaged over Phases 1--4 (Fig. 7c).  My fitting of the power-law plus 
annihilation spectrum to each VP where this point was visible to EGRET showed 
that the spectral anomaly apparently occurred in more than one event of 
amplitude $\sim 10^{-7}$ photon/(cm$^{2}$ s) on a characteristic timescale 
$\sim 100$ d.  This is a good example of how variability was clearly 
detected with a particular duty cycle, so I adopt its parameters as 
standard, for convenience.  They are close to the approximation that the
error varies inversely with the square root of the elapsed time, which is
true for Gaussian counting errors.  I conclude that the general upper limit 
on transient $p\overline{p}$ 
annihilation fluxes is $\sim 10^{-7}$ photon/(cm$^{2}$ s) for duty cycles 
$\sim 1/16$ of Phases 1--4 (i.e. 100 d), is scaled by a factor $\sim 10$
in the Galactic plane, and in general varies roughly as the reciprocal
of the square root of the time-scale of transient.

In terms of ETI activity at an unknown distance $R$ pc, the above numbers 
translate into the consumption of $\sim 1$--$5~R^{2}$ ton/s of antiprotons, or
$\sim 10^{8} R^{2}$ tons total.  The value of $R$ is completely undetermined; 
as an example, ETI activity within the Oort cloud at $\sim 0.25$ pc would 
be detectable if it involved $\sim 5 \times 10^{6}$ ton of antiprotons on 
the time-scales discussed above.  This is about 15 
orders of magnitude in excess of what 
could be expected to be produced on Earth in the near future [20].

My result may be compared with the predictions of earlier authors, taking 
into account the different combinations of antimatter mass and distance
$R$ which they used.  Thus Matloff and Mallove [21] considered a  rocket with 
payload mass $10^{10}$ kg, $\Delta v = 0.2 c$ and acceleration time 10 yr, 
which would burn  $\overline{p}$ at a rate of 1 kg/s.  At their reference 
distance of 10 pc it 
would be quite undetectable; EGRET would have detected it at $10^{12}$ km = 
6500 AU.  A more modest proposal for a crewed mission to $\alpha$ Centauri
(5000 ton payload on first leg of a 20-year return mission, accelerating at
$1~g$ to $0.245~c$ in 3 months,
[22]) burns a smaller amount of p- ($\simeq 15000$ tons) on a shorter 
time-scale and is 
detectable at a similar distance, 4500 AU.

My result may also be compared with the detectability radii for 
present-day human-planned interstellar spacecraft, i.e. the fusion-powered
Daedalus [23] and Orion [24] designs; by a variety of X-ray, gamma-ray, EUV 
and neutron detection techniques, these craft were found to be detectable
at ranges up to only $10^{9}$ km [10].  Presumably the antimatter propulsion 
mission to $\alpha$ Cen suggested by Forward (1 ton payload one-way,
180 kg $\overline{p}$ consumed in $< 1$ yr [11]), being much smaller, 
would be more difficult to detect by these conventional means.  However EGRET
would have detected it at a similar distance, $1.5 \times 10^{9}$ km =
10 AU.  In human terms, the missions proposed in [10] and [22]
can perhaps be thought of as the most primitive possible uncrewed and
crewed antimatter missions, respectively, to $\alpha$ Cen.

\subsection{Towards an Improved Measurement}

There are several approaches which might be explored in order to improve 
upon this result. An immediately practicable one would be to search for low 
duty cycle transients in the EGRET data.  As emphasised in \S 4.1 my search 
in the average Phase 1--4 data had some sensitivity to constant sources and 
those
of duty cycle $\ge 1/16$.  A search in the individual VP data might discover 
transients on much shorter time-scales.  This would involve performing the 
analysis steps 1--6 of \S 2.3 on all of the 170 individual VPs in Phases 
1--4.

Improvements in future high energy $\gamma$-ray telescopes would obviously 
lead to better results.  The problems reported in \S 3 with EGRET channels
0--2 show that sensitive and calibrated responses are a critical requirement at 
energies 30--100 MeV in order to determine the power-law background.  Better 
energy resolution overall would enable the shape of the feature, which I 
defined in terms of channels 4 and 5 and assumed to be due to $p\overline{p}$ 
annihilation,
to be determined more accurately.  A larger effective area would obviously 
increase the sensitivity of the search.  These improvements will be achieved by
the upcoming Gamma-Ray Large Area Space Telescope (GLAST 
[25], planned for 2006), 
with 15 times the sensitivity of EGRET and $\sim 5$ times its energy 
resolution.

\section{Conclusion}

I conclude that it is practicable to measure $p\overline{p}$ 
annihilation spectra 
separately from underlying power-law backgrounds if the appropriate 
selection criteria are applied.  By hypothesis, any source emitting this 
spectral feature must be artificial; it is extremely difficult to imagine any 
other possibility except new laws of physics (\S 1).  When applied to the 
EGRET data the method would have detected a steady $p\overline{p}$ annihilation
spectrum 
down to levels $\sim 2 \times 10^{-8}$ photon/(cm$^{2}$ s), 
and transients on 
time-scales ranging down to $\sim 100$ d at levels ranging up to $\sim 
10^{-7}$ photon/(cm$^{2}$ s),
outside the Galactic plane, both numbers being about a factor 10 higher in the 
plane.  Variable emission detected from the known extragalactic source QSO 
2206+650 is presumably not related to ETI activity.

These results, limited though they are, are the first ever obtained in this 
field.  They exclude the presence of "human-scale" antimatter-powered 
space probes (such as might be constructed by humans in this century [11])
within a radius of $\sim 10$ AU, and more ambitious human-like
crewed interstellar craft out to several thousand AU.
They will be greatly improved by future high energy $\gamma$-ray 
missions such as GLAST.

\acknowledgements

I would like to thank the EGRET PI team and Seth Digel in particular for 
creating an excellent archive.  I am grateful to Prof. Sherry Cady of
Portland State University for obtaining criticisms of a previous version
of this paper.
~\\

\begin{center}
{\bf References}
\end{center}

~ \\
1. A. Riotto and M. Trodden, "Recent progress in baryogenesis", {\em Annual 
Review of Nuclear and Particle Science\/}, 49, pp.35--75, 1999. \\
2. Ya. B. Zel'dovich and I. D. Novikov, I. D., {\em Relativistic 
Astrophysics, Vol. 
2. The Structure and Evolution of the Universe\/}, ed. G. Steigman, 
University of Chicago Press, Chicago, pp.163--170, 1983. \\
3. C. Baltay, P. Franzini,, G. L\"{u}tjens, J. C. Severiens, D. Tycko, and D.
Zanello, "Annihilation of antiprotons at rest in hydrogen. V. Multipion 
annihilations", {\em Phys. Rev.\/}, 145, pp.1103--1111, 1966. \\
4. G. Backenstoss et al., "Proton-antiproton annihilations at rest into 
$\pi^{0} \omega$,
$\pi^{0} \eta$, $\pi^{0} \gamma$, $\pi^{0}\pi^{0}$, and 
$\pi^{0} \eta\prime$", 
{\em Nuclear Physics\/}, B228, pp.424--438, 1986. \\
5. M. Yu. Khlopov, S. G. Rubin, and  A. S. Sakharov, "Possible origin of  
antimatter regions in the baryon dominated universe", {\em Phys. Rev.\/}, D62, 
083505, 2000. \\
6. Yu. A. Golubkov and M. Yu. Khlopov, "Antiprotons annihilation in the 
Galaxy as a source of diffuse gamma background", {\em Physics of the Atomic 
Nucleus"\/}, 64, pp.1821--1829, 2001. \\
7. R. R. Hillier, {\em Gamma ray astronomy\/}, Oxford University Press, Oxford,
p.13, 1984. \\
8. I. A. Crawford, "Interstellar travel: a review for astronomers", {\em 
Quarterly Journal of the Royal Astronomical Society\/}, 31, pp.377--400, 
1990. \\
9. M. J. Harris, "On the detectability of antimatter propulsion spacecraft",  
{\em Astrophysics and Space Science\/}, 123, 297--303, 1986. \\
10. D. R. J. Viewing, C. J. Horswell, and E. W. Palmer, "Detection of 
starships", {\em JBIS\/}, 30, pp.99--104, 1977. \\
11. R. L. Forward, "Antimatter propulsion", {\em JBIS\/}, 35, pp.391--395, 
1982. \\
12. M. J. Harris, "A search for linear alignments of gamma-ray burst sources", 
{\em JBIS\/}, 43, pp.551--555, 1990. \\
13. D. J. Thompson et al., "Calibration of the Energetic Gamma-Ray Experiment 
Telescope (EGRET) for the Compton Gamma-Ray Observatory", {\em Astrophys. J.
Suppl. Ser.\/}, 86, pp.629--656, 1993. \\
14. P. Sreekumar et al., "EGRET observations of the extragalactic gamma-ray 
emission", {\em Astrophys. J.\/}, 494, pp.523--534, 1998. \\
15. S. D. Hunter et al.," EGRET observations of the diffuse gamma-ray emission 
from the Galactic plane",  {\em Astrophys. J.\/}, 481, pp.205--240, 1997.  \\
16. A. W. Strong, I. V. Moskalenko, and O. Reimer, "Diffuse Galactic continuum 
gamma rays", in {\em The Fifth Compton Symposium\/}, ed. M. L. McConnell and 
J. M. Ryan, American Institute of Physics Proceedings 510, pp.283--290, 2000. \\
17. S, W. Digel, S. D. Hunter, I. V. Moskalenko, J. F. Ormes, and M. Pohl,
"The origin of cosmic rays and the diffuse Galactic gamma-ray emission",
in {\em Gamma 2001\/}, ed. S. Ritz, N. Gehrels, and C. R. Shrader,
American Institute of Physics Proceedings 587, pp.449--458, 2001.\\
18. D. Petry, "A first EGRET-UNID-related agenda for the next-generation 
Cherenkov telescopes", in {\em Proceedings, The Nature of Unidentified 
High-Energy 
Gamma-Ray Sources\/}, ed. A. Carraminana, O. Reimer and D. J. Thompson, Kluwer 
Academic, Dordrecht, pp.299--319, 2001. \\
19. R. C. Hartman et al., "The third EGRET catalog of high-energy gamma-ray 
sources", {\em Astrophys. J. Suppl. Ser.\/}, 123, pp.79--202, 1999. \\
20. G. R. Schmidt, H. P. Gerrish, J. J. Martin, G. A. Smith, and K. J. Meyer, 
"Antimatter requirements and energy costs for near-term propulsion 
applications", {\em Journal of Propulsion and Power\/}, 16, pp.923--928, 
2000. \\
21. G. L. Matloff and E. F. Mallove, "Alien starship detectability: bursters
and skidmarks", 39th Congress of the International Astronautical Federation, 
October 8-15 1988, Bangalore, Paper IAA-88-552, 1988. \\
22. B. N. Cassenti, "Design considerations for relativistic antimatter
rockets", {\em JBIS\/}, 35, pp.396--404, 1982. \\
23. A. Bond and A. R. Martin (eds), {\em Project DAEDALUS\/}, British 
Interplanetary Society, London, 1978. \\
24. J. C. Nance, "Nuclear Pulse Propulsion", {\em IEEE Transactions on Nuclear 
Science\/}, NS-10, pp.177--182, 1965. \\
25. P. F. Michelson, "The Gamma-ray Large Area Space Telescope mission: science 
opportunities", in {\em Gamma 2001\/}, ed S. Ritz, N. Gehrels, and C. R. 
Shrader, American Institute of Physics Proceedings 587, New York, pp.713--721,
2001. \\

\clearpage

\begin{table*}
\begin{center}
\begin{tabular}{lcccccccccc}

\tableline                          
                   
Channel number &  0 &  1  & 2 & 3 &  4 &  5 &   6  &  7 &   8 &   9 \\
\tableline

Lower edge [1] &  30 &  50 &  70 & 100 & 150 & 300 & 500 & 1000 &  2000  
& 4000 \\

Upper edge [1] &  50 &  70 &  100 & 150 & 300 &  500  & 1000 &  2000 &  4000 
& 10000 \\
\tableline

\end{tabular}

\end{center}

Note: \\
1. Energies in units of MeV \\

\caption{EGRET energy channels}

\end{table*}
~\\

\begin{table*}
\begin{center}
\begin{tabular}{lccccccc}

\tableline                          
                   
Zone  &  Step  & Step & Step &  Step & Step & Step & Flux, [2] \\
~[1] & 1 & 2 & 3 & 4 & 5 & 6 & photon/(cm$^{2} s$) \\

\tableline

$|b| > 8.5^{\circ}$ & 234270   & 8492 & 3456 & 387 & 12 & 1 &
$<2.3 \times 10^{-8}$ \\

$|b| \le 8.5^{\circ}$ & 24480 & 1639 & 138  & 138 & 29 & 0 & 
$<2.6 \times 10^{-7}$ \\
\tableline

\end{tabular}

\end{center}

Notes: \\

1. $|b| \le 8.5^{\circ}$ is the Galactic plane. \\
2. Flux 99\% confidence upper limit. \\
Other table entries represent the number of sources surviving each step in 
the selection procedure. \\

\caption{Search statistics, and results}

\end{table*}

\clearpage

\begin{figure}

\caption{ Normalized spectrum of photons resulting from "cold" antiproton 
annihilation; an analytic formula is given by [4].}

\caption{  Simplified diagram of EGRET.  Overall dimensions 1.65 m 
diameter $\times$ 2.25 m height.  Dashed line - incoming photon.  Full 
lines - $e^{-}$ and 
$e^{+}$ generated by pair production within Ta layer.  TOF = time of flight.}

\caption{  The $p\overline{p}$ 
annihilation spectrum of Fig. 1 binned into the EGRET energy 
channels from Table 1.}

\caption{ Schematic illustration of the spectrum resulting from a 
$p\overline{p}$ 
annihilation signal (Fig. 1) superimposed on a power-law background.  Dashed 
lines - individual components.  Full line - sum of the two components.  Numbers
at top of figure represent the weighted mean energies of the EGRET channels.  
For ease of illustration, the amplitude of the annihilation spectrum is much 
greater than a typical case from my selected spectra.}

\caption{ Procedure for accumulating EGRET spectra for points away from the 
Galactic plane (schematic).  Intensity at a point A = average intensity 
measured over all points within the approximated EGRET PSF (full circle, radius
$2.5^{\circ}$).  Counts at point A = sum of all counts registered within the 
circle.  
The procedure is repeated at every grid point (dashed circles centred at B, 
C...).}

\caption{ Histogram - distribution of 525 measured 
$p\overline{p}$ annihilation amplitudes in 
terms of their statistical significances (amplitude divided by statistical 
error).  Full line - part of the expected distribution for measurements 
randomly (normally) distributed around a null result.  Note the positive 
offset of the measured distribution and the excess of measurements above the 
full line for significances $> 2.5 \sigma$.}

\caption{ (a) A typical EGRET spectrum, showing weak or no 
$p\overline{p}$ annihilation feature.  Phase 1--4 spectrum, $l = 
117.25^{\circ}$, $b = 11.75^{\circ}$.  Data
points - intensities in EGRET channels 0-8.  Full line - spectral model fit 
with power-law plus annihilation function (Figs. 1, 4) of amplitude $0.29 \pm
1.08 \times 10^{-8}$ 
photon/(cm$^{2}$ s). (b) Spectrum showing a spurious annihilation 
feature (full line) due 
to a systematic error in channel 0, such as is responsible for most
of the measurements with significance $> 2.5 \sigma$ in Fig. 6.  Phase 1--4 
spectrum, $l = 197.25^{\circ}$, $b = 25.25^{\circ}$, $p\overline{p}$ amplitude 
$2.88 \pm 0.45 \times 10^{-8}$ 
photon/(cm$^{2}$ s).  Dashed line --- power law fit to channels 1--8.  
(c) Phase 1--4 spectrum of candidate object $l = 
107.25^{\circ}$, $b = 11.5 ^{\circ}$, 
rejected due to identification with QSO 2206+650.  Annihilation 
amplitude $5.05 \pm 1.19 \times 10^{-8}$ 
photon/(cm$^{2}$ s). (d) Variability of flux 
from $l = 107.25^{\circ}$, $b = 11.5^{\circ}$, from measurements during each VP
(top label) when the point was within the EGRET aperture.  Dashed line --- 
overall amplitude $5.05 \times 10^{-8}$ 
photon/(cm$^{2}$ s) during Phases 1--4 from {\em (c)\/}.  Mission day 0 = 
16 May 1991.}

\end{figure}

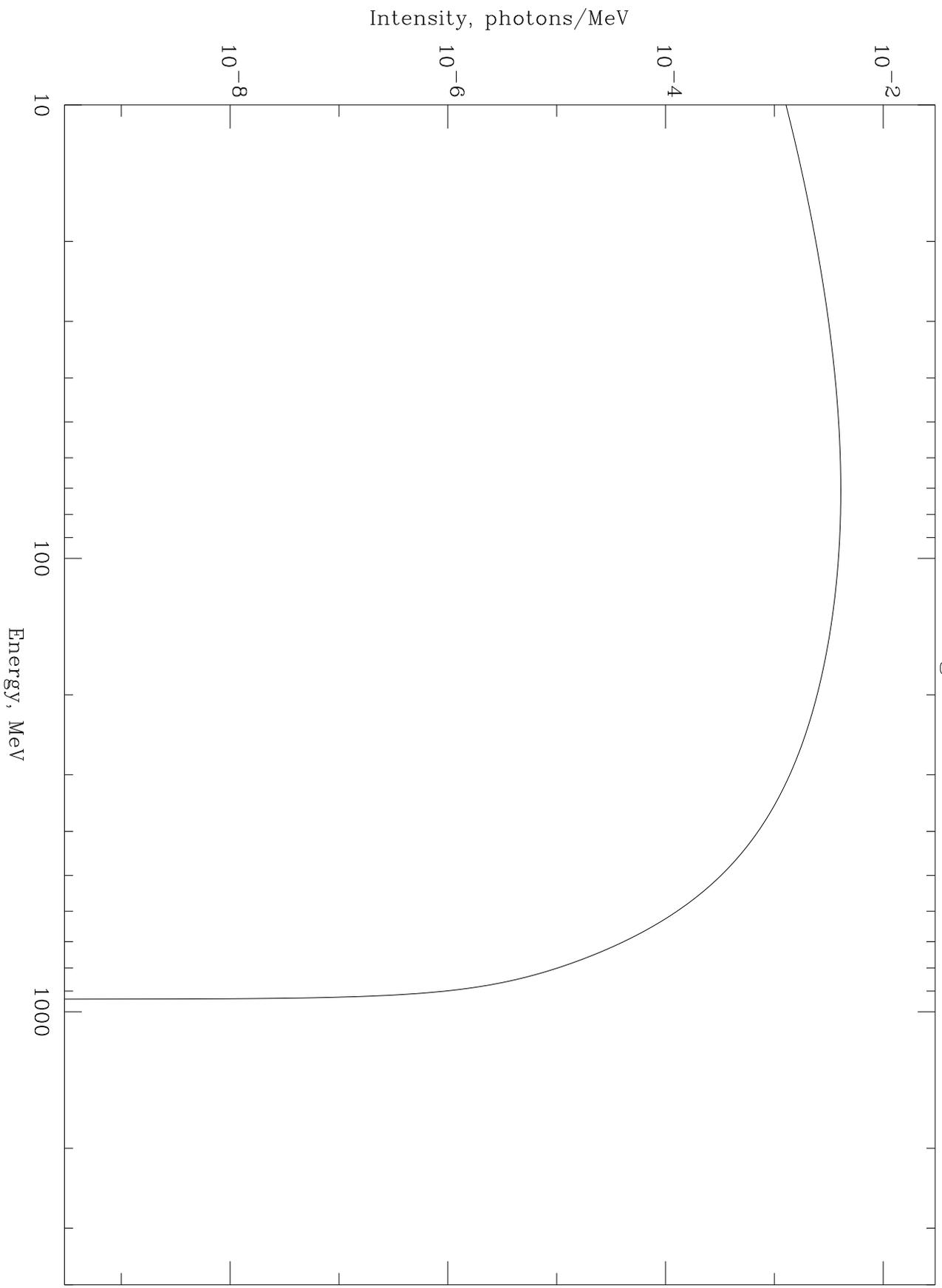
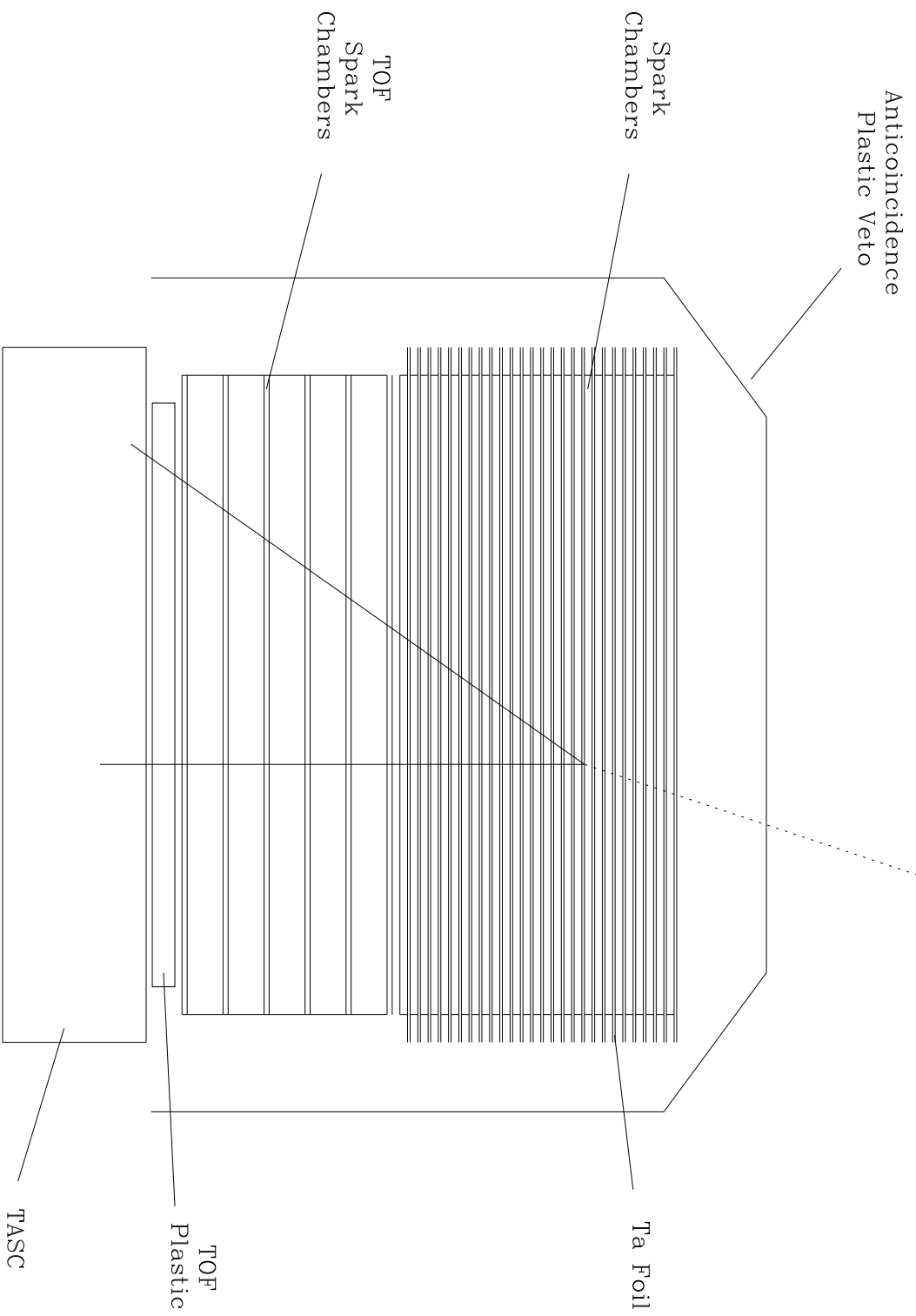
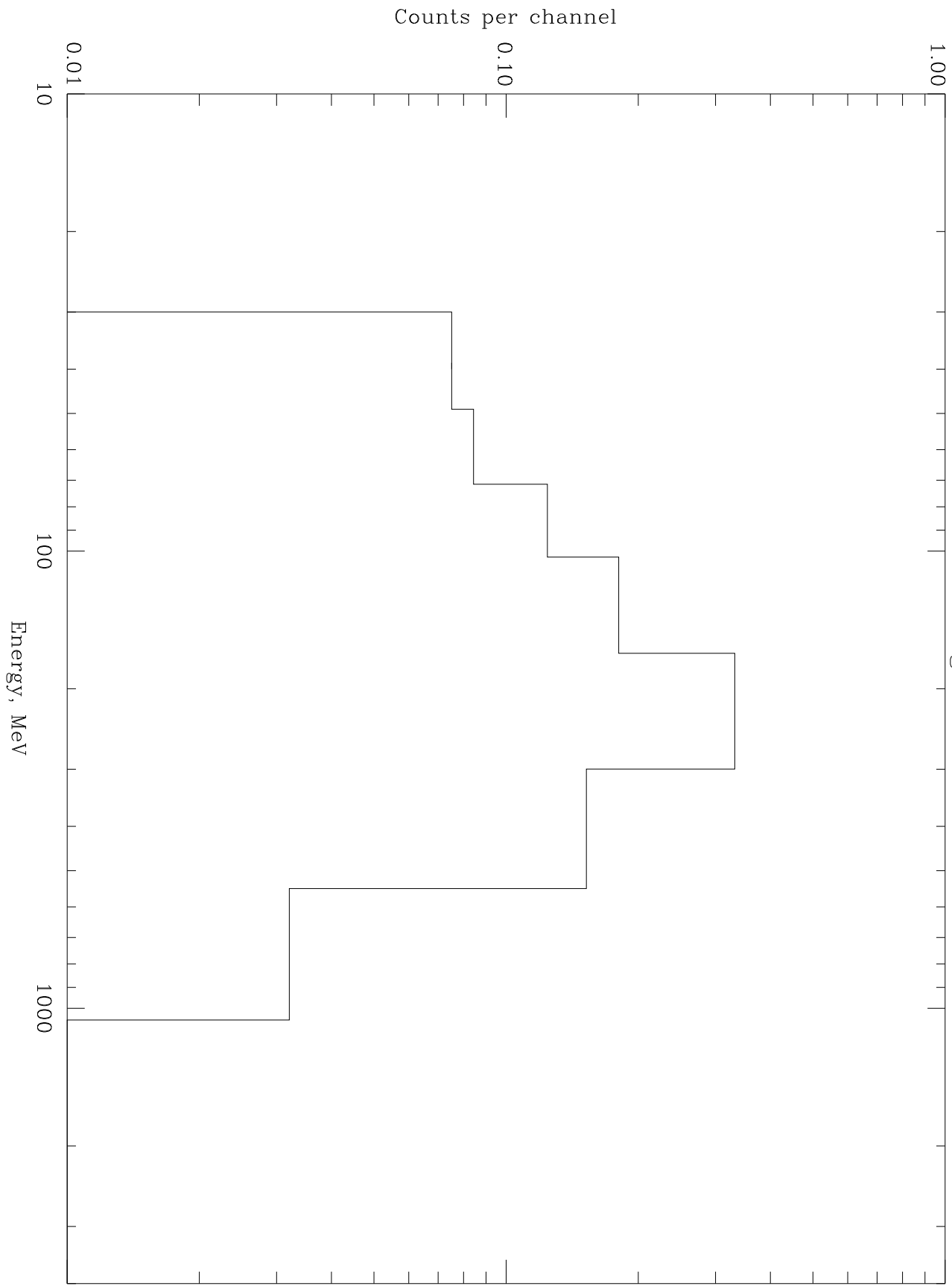
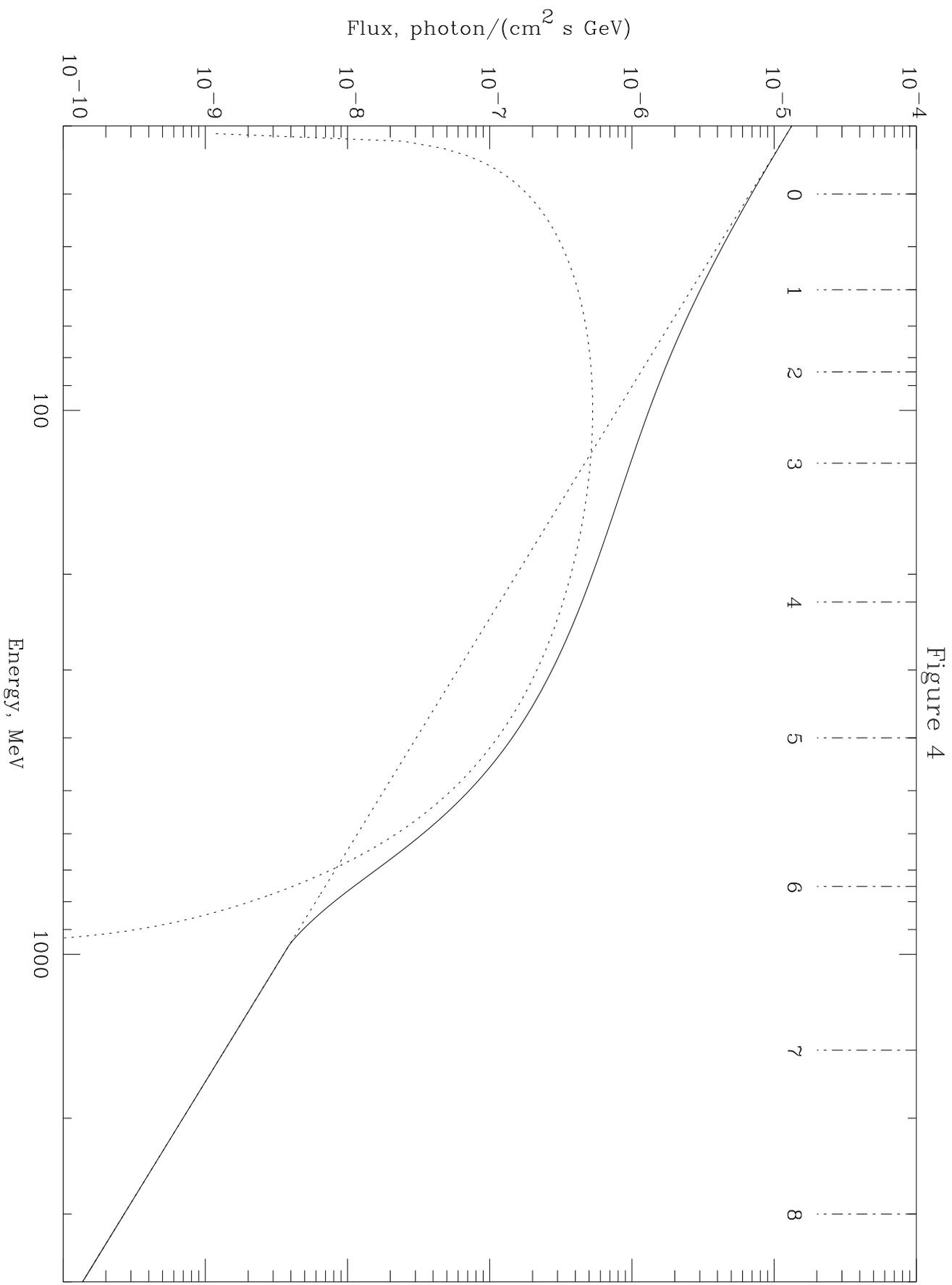
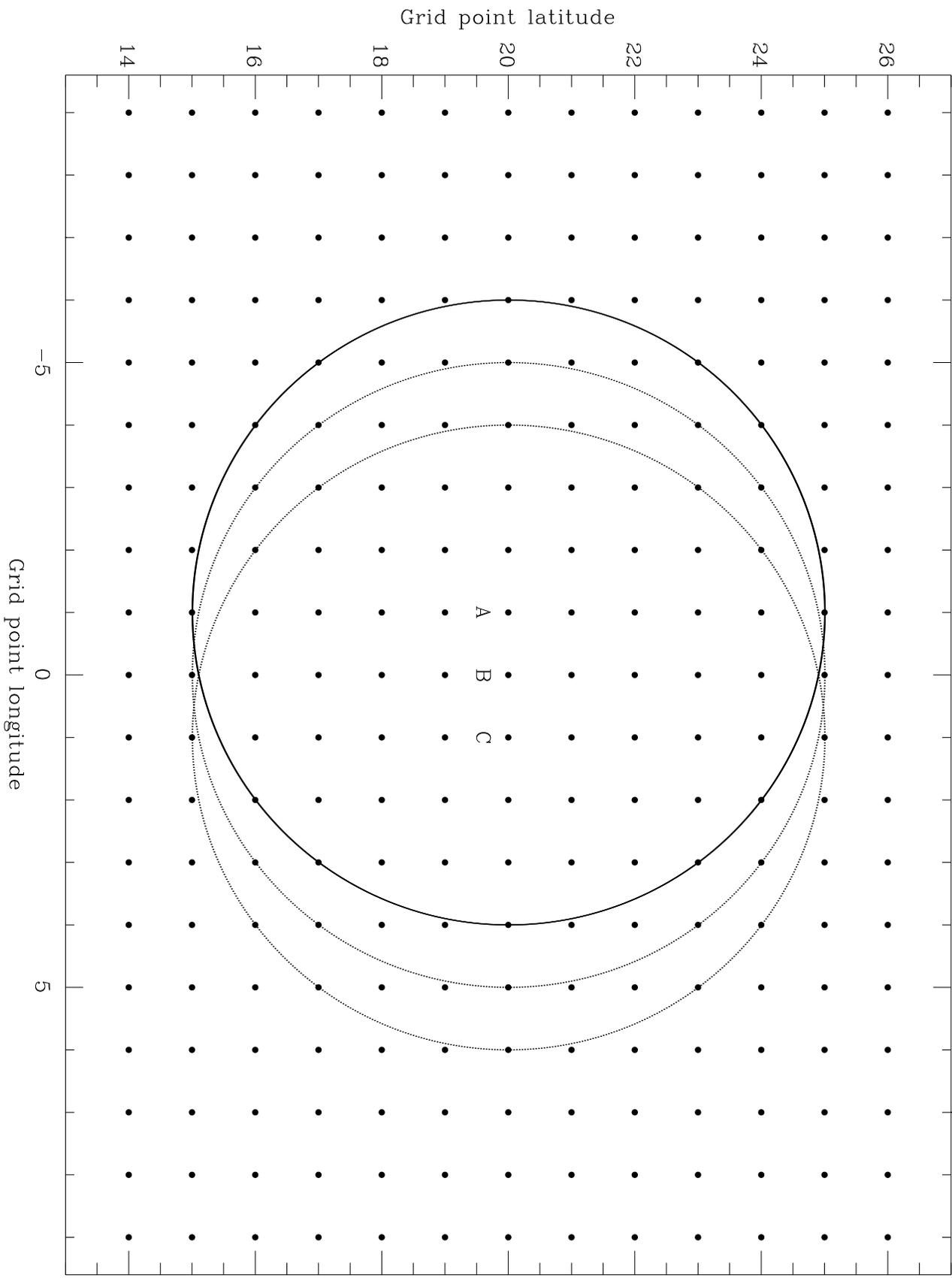
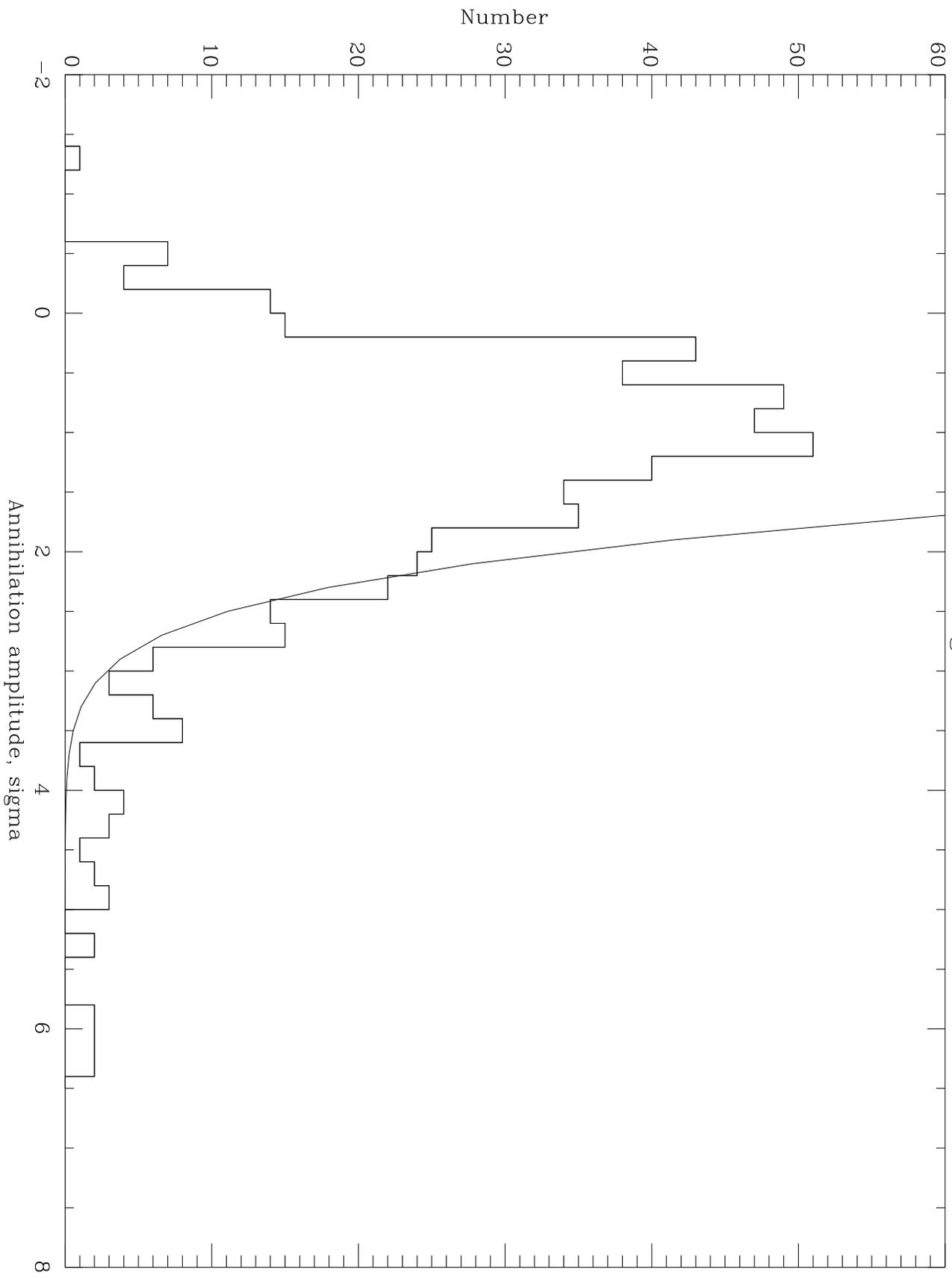
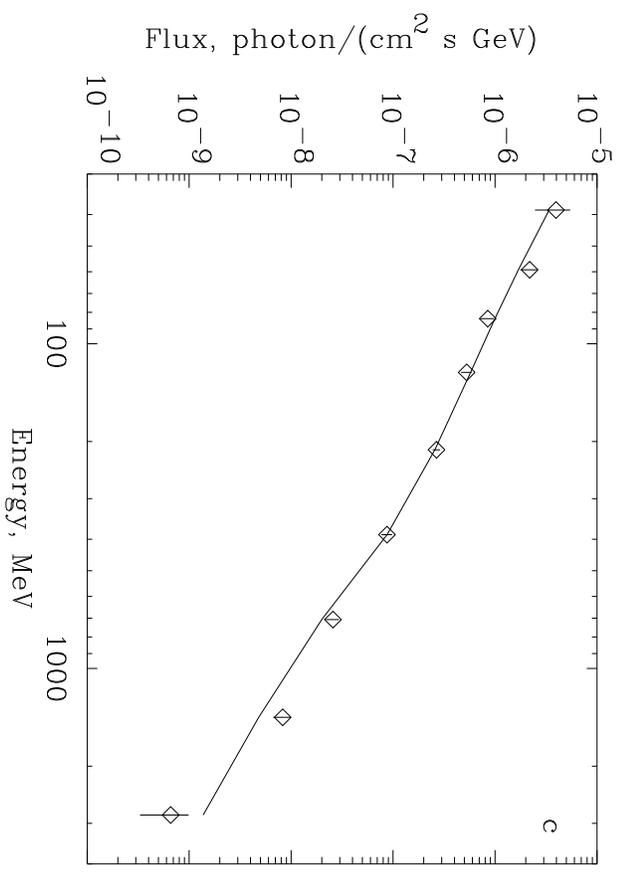
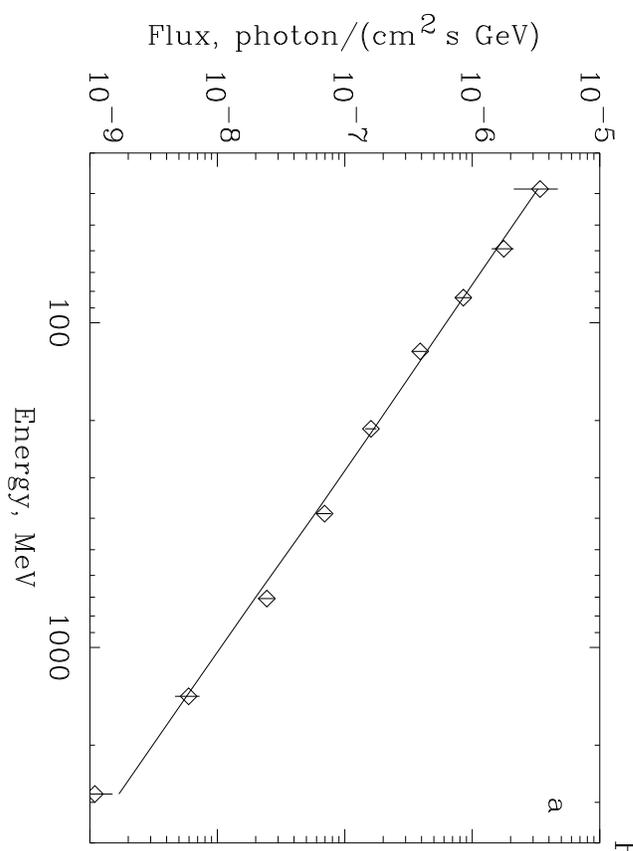
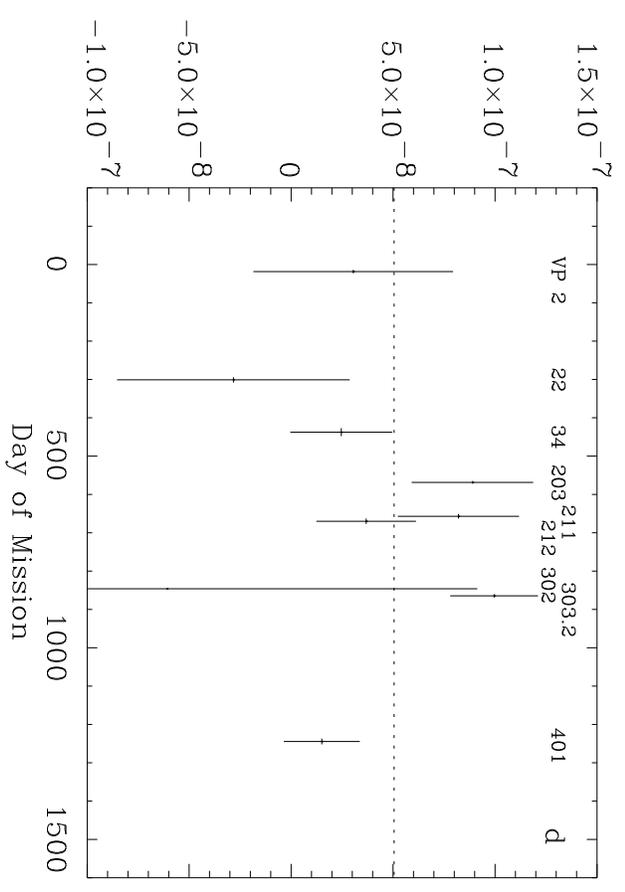
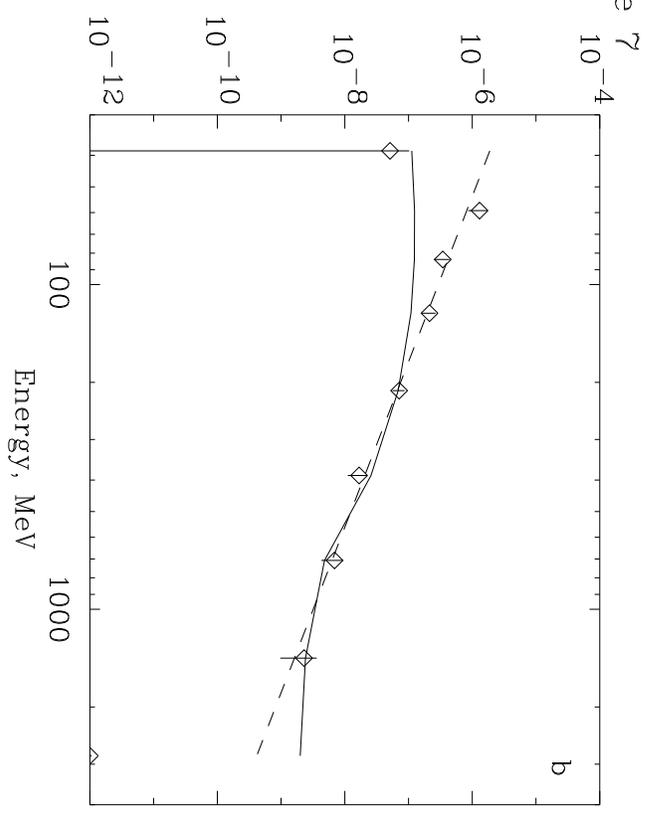
\end{document}